# Modelling and Implementation of ITWS: An ultimate solution to ITS

Nirmalendu Bikas Sinha[1], Manish Sonal[1] , Makar Chand Snai[1],  R.Bera[2] ,And M.Mitra[3]

**Abstract**— Casualties due to traffic accidents are increasing day by day. Think of this message being displayed on your computer screen while you were driving "there's a possibility of collision with a car in the next few minutes if you go on driving with this speed and direction". Our research is intended towards developing collision avoidance architecture for the latest Intelligent Transport System. The exchange of safety messages among vehicles and with infrastructure devices poses major challenges. Specially, safety messages have to be adaptively distributed within a certain range of a basically unbounded system. These messages are to be well coordinated and processed via different algorithms. The purpose of the paper is to discuss the ITWS (intelligent transportation warning system) ,we have discussed the Assisted Global Positioning System(AGPS) system providing additional positioning information at variable conditions. We have also discussed study the Data fusion and kalaman filter in details. The performance of kalman filter and output are discussed. Hardware realization of this model is achieved through software defined radio (SDR).

———————————— ◆ ————————————

## 1. INTRODUCTION

Intelligent transport systems (ITS) aims to apply information technology, communications technology, and sensor technology, including the internet to transportation systems to improve travel safety, reliability, and convenience, increase mobility, mitigate traffic congestion, and reduce fuel consumption and emissions. The targets of ITS development include a myriad of products and services such as internodal transportation systems, intelligent traffic control systems, in-vehicle technologies, safety enhancement technologies, traveller advisor system, and so on.

Intelligent transport systems show great potential in improving existing transportation systems.To implements ITS, intelligent transportation warning system (ITWS) functional architecture has been considered. ITWS architecture used earlier involved GPS and conventional RADAR to obtain sensorial fusion. These system have limitation that GPS systems mainly depend upon weather conditions, moreover they are not reliable in different environments such as inside buildings, under dense foliage, and in urban canyons because there aren't enough number of satellites in view. Considering these difficulties we have tried to use Hybrid communication system and MIMO RADAR system to obtain sensorial fusion using kalman filter.

MIMO Radar can drastically change the system Performance of our proposed collision avoidance architecture as a remote sensing equipment, more ever the use of MIMO radar (with colocated antennas) can exploits the advantage like increased diversity of the target information, excellent interference rejection capability, improved parameter identifiability[1]-[4], and enhanced flexibility for transmit beam pattern design .It also can help for the target tracking, target localization. The degree of freedom introduced by MIMO radar improves the performance of the radar systems in many different aspects. Use of OFDM in this Doppler radar with colocated antennas (OFDM-MIMO radar) can again further improve the system gain in terms of target estimation performance. But for security purpose DSSS have an excellent advantage, use of this along with OFDM can give a remarkable advantage in collision avoidance architecture.

The paper is presented in the following manner; first our proposed collision architecture is presented involving the Digital MIMO Radar and Hybrid communication. Then the Data fusion and kalaman filter have been developed and discussed in details with its output and structure analysis.

————————————————
- *Prof. Nirmalendu Bikas  Sinha, corresponding author is with the Department of ECE and EIE , College of Engineering & Management, Kolaghat, K.T.P.P Township, Purba- Medinipur, 721171, W.B., India.*
- *Manish sonal is with the Department of ECE, College of Engineering & Management, Kolaghat, K.T.P.P Township, Purba- Medinipur, 721171, W.B., India.*
- *Makar Chand Snai is with the Department of ECE, College of Engineering & Management, Kolaghat, K.T.P.P Township, Purba- Medinipur, 721171, W.B., India.*
- *Dr. R. Bera is with the S.M.I.T, SikkimManipal University, Majitar, Rangpo, East Sikkim, 73713.*
- *Dr. M.Mitra  is With the Bengal Engineering and science University, Shibpur, Howrah, India .*

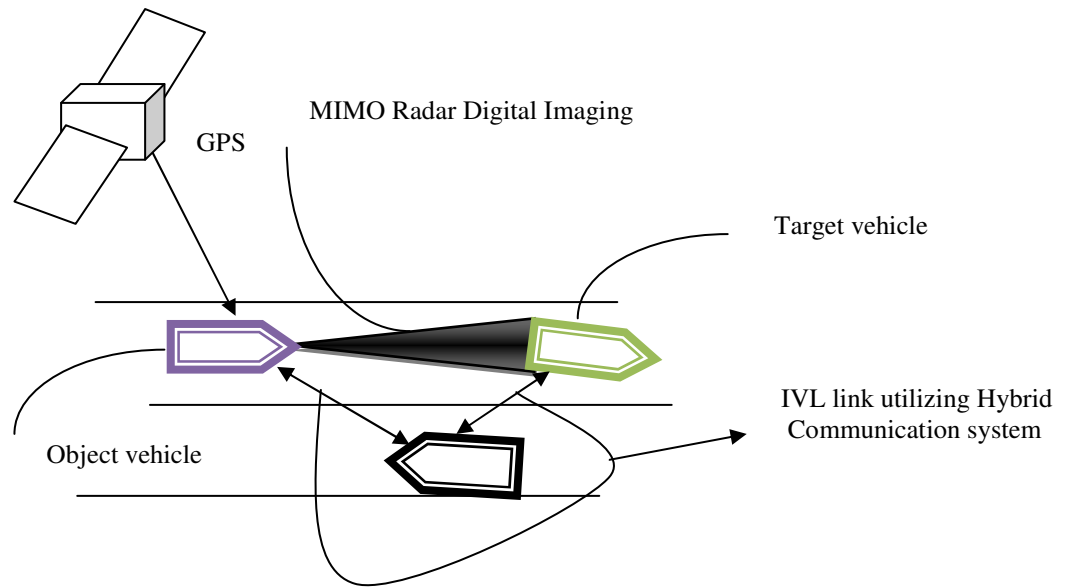

Fig.1.Concept of collision avoidance architecture for ITS

Fig.1 illustrates the MIMO Radar digital imaging, GPS and IVL Link incorporated with Hybrid Communication system would give an ultimate collision avoidance system for ITS. The object vehicle is getting information about its location from the GPS system; it is comparing its position with the data broadcasted by other vehicles through IVL. To get information about the nearby vehicles and pedestrians the object vehicle would analyse the scatterer using MIMO radar .Radar system would accurately tell about the type and position of target vehicle in the best efficient way possible. Along with the information provided by the MIMO Radar about spectrum the direction of the target vehicle can be well defined. On collecting the in formations about the location, type and accurate speed of the target vehicle, the object vehicle would run collision algorithm as per our architecture to get the proper speed, driving line and enough information's to avoid collision.

## 1.1 ITWS architecture

Fig.2 gives the basic ITWS architecture, each of its blocks are explained below.

*Assisted GPS (A-GPS):*
The assisted GPS is the idea to exploit the availability of the bi-directional link to transfer information to augment the functions of mobile phone that contains a partial or full GPS receiver. A-GPS consists of three parts: location server, mobile    station (MS) with partial or full GPS receiver, cellular wireless communication link which is comprised of Mobile switching centre (MSC) and Base switching centre (BSC) mainly. AGPS technique is one way to characterize targets for identification**.** The information transferred are obtained from the wireless infrastructure, wireless handset, and from a location server. This information is called assistance information, which is used by the wireless A-GPS. Among the two type of assisted GPS system [5] (MS-assisted GPS, MS-based GPS) the MS-assisted GPS are widely used, here MS provides assistance data firstly to BSC then it goes to MSC.The local server extracts this assisted data from MSC and compute the particular user's rough position. The main function of location server is to monitor satellites in the sky and compute the exact position of user by combining the data from users and data from the satellite, after computation the location information is transmitted to the respective user. Fig. 3 shows this basic architecture. Wireless cellular communication link is just to provide a data exchange channel for the users and location server.

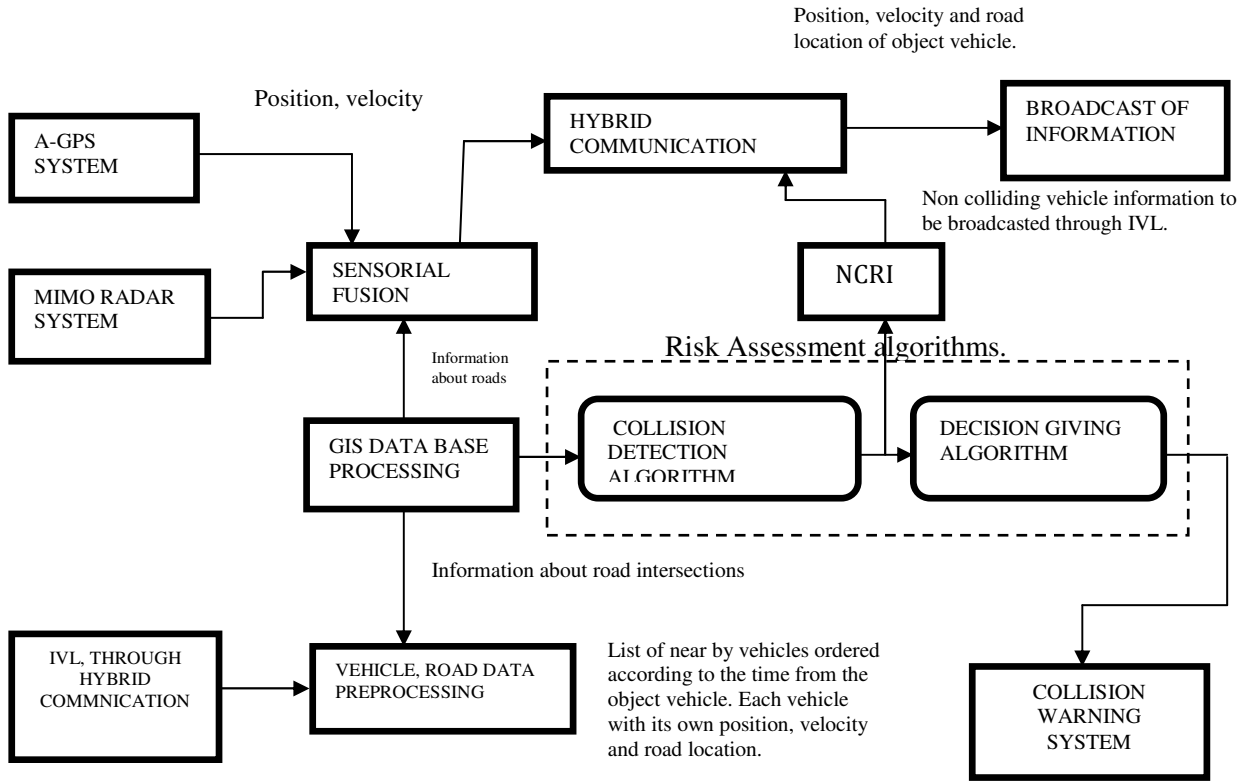

Fig 2.  ITWS (Intelligent Transportation Warning System) architecture.

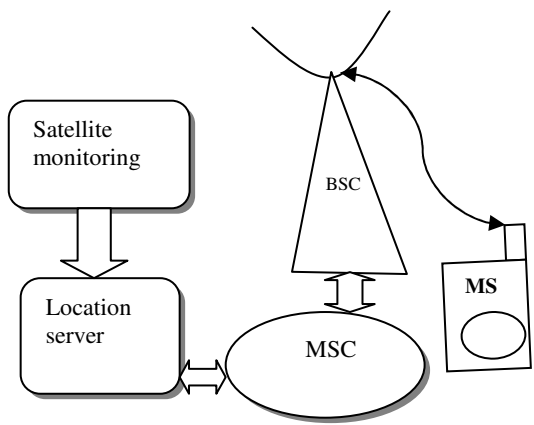

Fig 3.Block diagram of GPS system

*IVL:* Allows the information transmission among vehicles. This information is used to calculate the movement relative to the others vehicles.

*Vehicle Road Data Pre Processing*:

The basic information like velocity, distance(in time),position and road location of all nearby vehicles are pre processed before being executed by the collision detection algorithm. The above said information are collected through IVL (inter vehicle link) and GIS(Global information system) data base.

*GIS Database Processing:*

This is the GIS management algorithms to get the adequate localization information. The information must include: road, kilometric point, critical points near vehicle (roads intersections, roundabouts, etc).All this management must be done in quasi real time.

*Sensorial Fusion:*

Integrates all measurements from the sensors and GIS database .Thus it obtains reliable, dynamically variables and appropriate vehicles information in order to assess the collision risk.

*Processing Algorithm:*

(a) *Collision Detection Algorithm:*

They are the core part of the system and are related to artificial intelligence. Using the information from intra vehicle link, sensorial fusion system and trajectory prediction algorithms, the system has to assess the collision risk.

(b) Decision Giving Algorithm:

Collision warning interface with the driver. Based on the results from collision risk assessment this component has to produce adequate warnings to the driver. These warnings have to be carefully showed to driver in order to avoid dangerous driver's behaviours due to incorrect information.

Non Collision Risk Information (NCRI):

The actual result given by decision algorithm passes through several filters. The car or objects in a nearby radius/risk range are eliminated after filtering. The NCRI block would transmit the information via IVL to all the nearby objects so that the decision taken by other vehicles become more robust and optimal.

Broadcast of Information:

Position, velocity, location of the object vehicle along with the NRCI is being broadcasted to the nearby object through IVL utilising Hybrid Communication.

Collision Warning System:

For the generation of messages, following parameters are considered: Probability of collision and time of possible collision. According to these two parameters the warning messages are classified in three categories: High, medium and low risk. Finally, a list with the classified warning messages is produced. Additionally, the direction of the risk is calculated using the target vehicles' position and velocity data. The objective car is divided in eight zones of possible collision: front, front-right, right, rear-right, rear, rear-left, left and front-left. This information is included into the Warning message.

*Digital Radar System*

Radar (Fig.4) systems have been used extensively over the past decade for a variety of applications and in a multitude of configurations. One of the relatively more modern implementations—although its origins are traced back to the end of the 1950—is imaging radar, in particular, synthetic aperture radar (SAR).Imaging radars are used to obtain visual in- formation about the environment of interest, often with the goal of discerning particular objects concealed either intentionally or un- intentionally in the background. These radars can be geared toward certain scenarios, such as discovery of buried mines and unexploded ordnance, or assessment of polar ice cap dynamic, or as a surveillance and target tracking tool in reconnaissance operations.

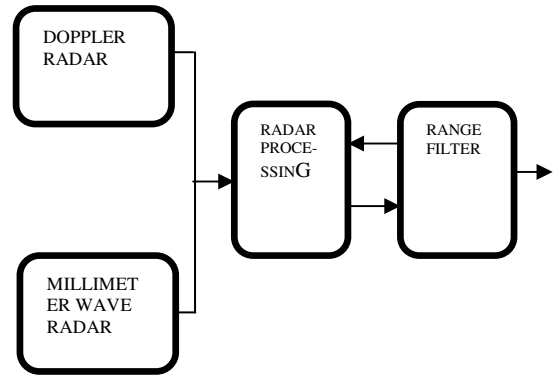

Fig.4.Digital Radar System

The MIMO systems have gained popularity and attracted attention of late for their ability to enhance all areas of system performance. Inspired by the success of MIMO systems in communications [6][7] , several publications have advocated the concept of MIMO Radar [1-3] from the system implementation point of view [4], as well as for processing techniques for target detection and parameter estimation[3] .Target parameters of interest in radar systems include target strength, location, and Doppler characteristics.

MIMO radar system is one of the vital parts of the proposed Collision avoidance Architecture. It is used to obtain sensorial fusion along with MIMO communication system in order to get detailed information of the objects that are approaching it. Compared with conventional array radar, MIMO radar has a lot of advantages. MIMO radars can increase the number of available degrees of freedom, which can be exploited to improve resolution, clutter mitigation, and classification performance. Exploiting the independence between signals at the array, MIMO radars capitalize on RCS scintillations with respect to target aspect to improve the radars performance. MIMO radar architecture Fig.5 employs multiple transmit waveforms and has the ability to jointly process signals received at multiple antennas created by controlling correlations among transmitted waveforms .

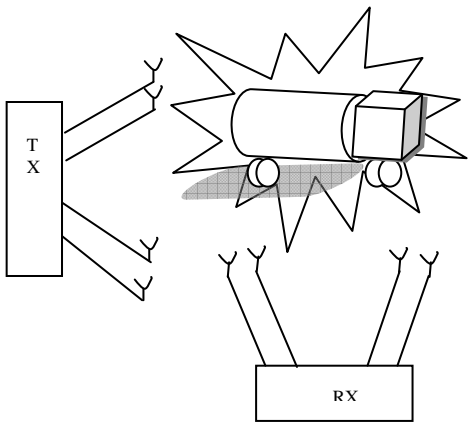

Fig.5.MIMO Radar System

Hybrid communication system:

The information about the speed, location, position and collision probability that would be broadcasted by vehicles to other nearby objects would be through our proposed Hybrid communication system. This system is basically a combination of Orthogonal Frequency Division Multiplexing (OFDM), MIMO System, MIMO OFDM Systems, DSSS system technologies supporting very high data rate for static and mobile users. Data rate can reach up to 100Mbps for Mobile users. More over the data transmission is more secure and robust. Following technologies comprises to Hybrid communication system.

# CALM System Architecture (21210) (Rev. Geneva)

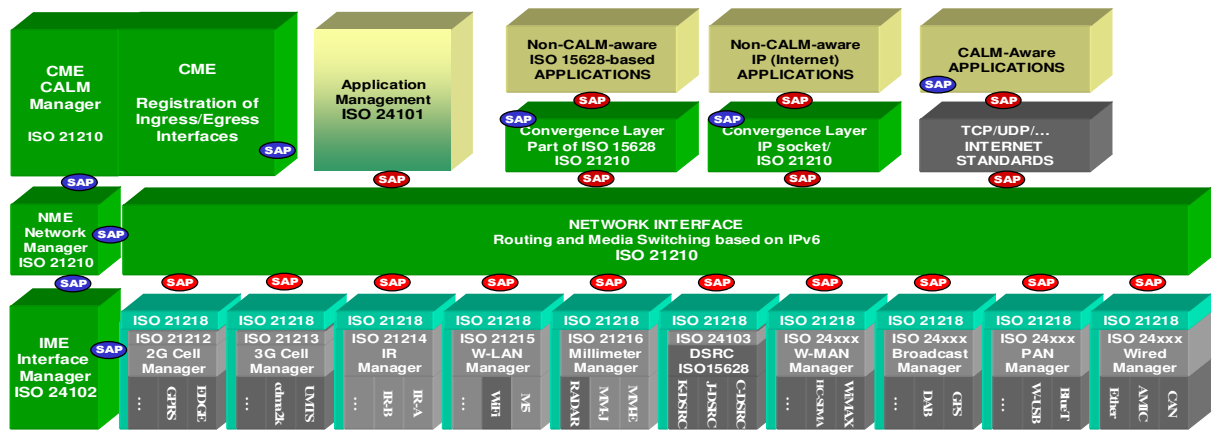

Fig.6. CALM [Continuous Air interface for Long and Medium distance] System Architecture.

CALM [8], continuos communication for vehicles, is a new World Standard for ITS operation and is depicted in above Fig.6. It includes Millimeter wave radar, GPS, 2G air interface to support ITS activitits.

## 2. 4G AND ITS CONVERGENCE

The Continuous Communication supported by 4G can be best exploited in ITS application. Every car should announce its location continuously through Inter Vehicle Link IVL as depicted in Fig.1. The others nearby cars will listen it and get the position of the nearby car and thus generate its own Automatic Cruise Control ACC. The position of the car can be estimated through GPS. But the errors in GPS can be corrected with the use of Assitance from Mobile tower and local vehicular radar installed in the car.

## 3. CHOICE OF TECHNOLOGY FOR ITS

There are a few possibilities concerning the communication between the road infrastructure and the vehicles as well as inter car communication. These are both subject to similar constraints and limitations. For ITS application the communication will have to be established and completed within a very short period of time at high speed. This is a limitation concern only short range technologies. The use of long range technologies like 3G would eliminate those issues [9].

## 4. REMOTE SENSING TECHNOLOGIES USED IN ITS:

A number of remote sensing devices are used for ITS application. The commonly known remote sensing devices based on using sensors are pneumatic road tubes, inductive loops, magnetic sensors, piezoelectric sensors, video cameras, infrared lasers sensors, microwave (MW) radars, and ultrasonic sensors [10]–[14]. Any kind of sensor will provide a specific mechanism of detecting and classifying vehicles and has its own advantages and disadvantages. Since user needs and classification conditions can differ, no sensors and corresponding techniques have proven to be the best for all possible applications [15][10].Therefore, any new classification technique providing specific advantages can be of great interest for the highway agencies. In certain situations, some benefits can be provided by MW radar sensors. MW sensors do not require installation in the roadway, thus making sensor calibration and maintenance easier and less disruptive. This is important for use in high-volume urban freeways, highways, and other locations where access to the roadway is extremely limited and expensive. Another strength is that MW radar sensors are largely immune to adverse weather and light conditions or vibrations. Such properties have lead to intense practical interest in MW classification systems [10][15]. In various traffic management applications, roadside mounted and forward-looking frequency-modulated continuous wave (FMCW) and noise-correlation radar units combined with continuous-wave (CW) Doppler sensors are commonly used [10], [16], and [17]. Both of these MW radar sensors are primarily intended for extraction of vehicle length and shape information. However, high resolution in the distance domain is required to obtain accurate vehicle shape information from a roadside mounted sensor. As a result, the FMCW radar classification system considered in [16] gives only 75% accuracy when traffic into five categories. A forward-scattering CW Doppler radar was used to obtain a vehicle signature in [18]. The vehicle signatures obtained using Doppler radar was also used in [19] and [20]. Because of the high variability in geometric shapes of vehicles, extracting sufficient information from the signatures for the detailed classification of vehicles is a difficult task. Therefore, such types of classification systems are feasible for categorization into a relatively small number of vehicle classes such as small, medium, and large cars [18], [19] or tracked and wheeled vehicles [20]. To enhance classification accuracy and increase the number of vehicle types being classified, a vehicle classification system using down-looking spread spectrum MW radar was proposed in [21]. In this system, the sensor was mounted above the roadway in such a way that vehicles pass directly below the sensor. Such installation made it possible to use the vehicle height profile as the main feature for classification. As a result, the reported classification accuracy was 99% for five vehicle types and a data set of 1706 vehicles.

## 5. GPS in ITWS

Global Positioning System is a Satellite-based system that uses a constellation of 24 satellites to give an accurate position of user and GPS provides a global absolute positioning capability with respect to a consistent terrestrial reference frame and considered as an absolute global geodetic positioning system. GPS receivers have been developed to observe signals transmitted by the satellites and achieve sub-meter accuracy in point positioning and a few centimeters in relative positioning. The GPS satellites are positioned in such a way that at least five to eight satellites are accessible at any point on earth and at any time. GPS is based on a system of coordinates called the World Geodetic System 1984 (WGS-84 whose coordinates are the latitude, longitude, and height) (Kaplan.E.D, 1996) and (Parkinson, 1996).GPS data is observed in WGS 84 and Universal Transverse Mercator (UTM). The most common map projection and grid system used for land navigation is the Universal Transeverse Mercator (UTM) system. A key advantage of the UTM projection is its preservation of the shape of the small areas on a map and its grid coordinates permit easy calculations using plane trigonometry (Langely B Richard, 2000). In UTM, the ellipsoid is portioned in to 60 zones with a width of $6^0$ longitude each. A scale factor of 0.996 is applied to the central meridian (Leick, 1995). The scale factor is to avoid fairly large distortions in the outer areas of zone. There are several sources of error that degrade the GPS position from few meters to tens of meters (Pratap Misra, 2001). These error sources are Ionospheric, Atmospheric delays, Satellite and Receiver Clock Errors.Multipath, Dilution of Precision, Selective Availability (S/A) and Anti Spoofing (A-S) as described by Hoffmann - Wellenhof et al (1998). The errors could be transmitted via VHF/UHF links and the users can make use of the corrections to fix their positions more accurately. These errors can be reduced to arrive at a more accurate estimate of coordinates of user by means of a recursive algorithm- kalman filter. The emphasis is given on the above errors to analyze the kalman filter (Grewel.M. S et al, 2001). The GPS receiver is giving both WGS-84 data observed in latitude, longitude & altitude and UTM data observed in Northing & Easting (Langely.R.B, 2000). In the conversion process of WGS 84 to UTM, accuracy must be obtained without distortions .In the work attempted by Ravindhra .et al (Feb, 2002), only the datum conversion from WGS- 84 to UTM and inaccuracy were discussed. The role of the noise in GPS is only at satellite and receiver segments.

## 6. RADAR in ITWS

Synthetic Aperture Radar (SAR) is a technique which uses signal processing to improve the resolution beyond the limitation of physical antenna aperture. In SAR, forward motion of actual antenna is used to 'synthesize' a very long antenna. SAR allows the possibility of using longer wavelengths and still achieving good resolution with antenna structures of reasonable size. The use of SAR for remote sensing is particularly suited for tropical countries. By proper selection of operating frequency, the microwave signal can penetrate clouds, haze, rain and fog and precipitation with very little attenuation, thus allowing operation in unfavourable weather conditions that preclude the use of visible/infrared system. Since SAR is an active sensor, which provides its own source of illumination, it can therefore operate day or night; able to illuminate with variable look angle and can select wide area coverage. In addition, the topography change can be derived from phase difference between measurements using radar interferometer. SAR has been shown to be very useful over a wide range of applications, including sea and ice monitoring, mining, oil pollution monitoring, oceanography, snow monitoring, classification of earth terrain etc. The potential of SAR in a diverse range of application led to the development of a number of airborne and space borne SAR systems. Using short unmodulated pulses like in this contribution leads to a range resolution which is proportional to the pulse duration. In the case of rectangular pulses the pulse duration is reciprocal the signal bandwidth. A precise analysis shows that the essential criterion for the achievable resolution is not the duration but the bandwidth of the processed signal if the possibility of pulse compression is exploited. Today, radar systems commonly use long pulses with large bandwidth because they offer high resolution on one hand and a convenient power ratio. The distance which has been covered by a reflected radar signal can be determined very precise except for a multiple of its wavelength. Thus, the range changes caused by changes of the geometry can be measured using periodical pulse repetition. The accuracy of this measurement is better the smaller the signal wavelength is. A pulse generation unit creates pulses with a bandwidth according to the aspired range resolution. They will be amplified by the sender and are transferred to the antenna via a circulator. The receiver gets the antenna output signal (echoes of the scene) amplifies them to an appropriate level and applies a band pass filter. After the demodulation and A/D conversion of the signals the SAR processor starts to calculate the SAR image. Additional motion information will be provided by a motion measurement system. A radar control unit arranges the operation sequence, particularly the time schedule. An inverse Fourier transform leads to the SAR image of the point target. (The amplitude term is the wave number representation of the point spread function and the linear phase term provides the correct location after the transformation.)

## 7. DATA FUSION AND KALMAN FILTER IN ITWS

The Multi-Sensor Data Fusion (MSDF) approach is described as the acquisition, processing, and synergistic combination of information gathered by various knowledge sources and sensors to provide a better understanding of a phenomenon under consideration Different MSDF techniques have been explored recently. These techniques vary from those based on well established Kalman filtering methods, to those based on recent ideas from soft computing technology. However, little work has been done in exploring architectures that consider the combination of both these approaches. In this work a novel MSDF architecture that combines these approaches is explored. This architecture is built integrating the fuzzy logic-based adaptive kalman filter developed recently by Escamilla and Mort and a fuzzy logic performance assessment scheme. The general idea explored in this approach is the combination of the advantages that both Kalman filtering and fuzzy logic techniques have. On the one band, Kalman filtering is recognized as one of the most powerful traditional techniques of estimation: the Kalman filter provides an unbiased and optimal estimate of a state vector in the sense of minimum error variance . On the other hand, the main advantages derived from the use of fuzzy logic techniques, with respect to traditional schemes, are the simplicity of the approach, the capability of fuzzy system to deal with imprecise information, and the possibility of including heuristic knowledge about the phenomenon under consideration.

## 8. Flowchart of DATA FUSION:

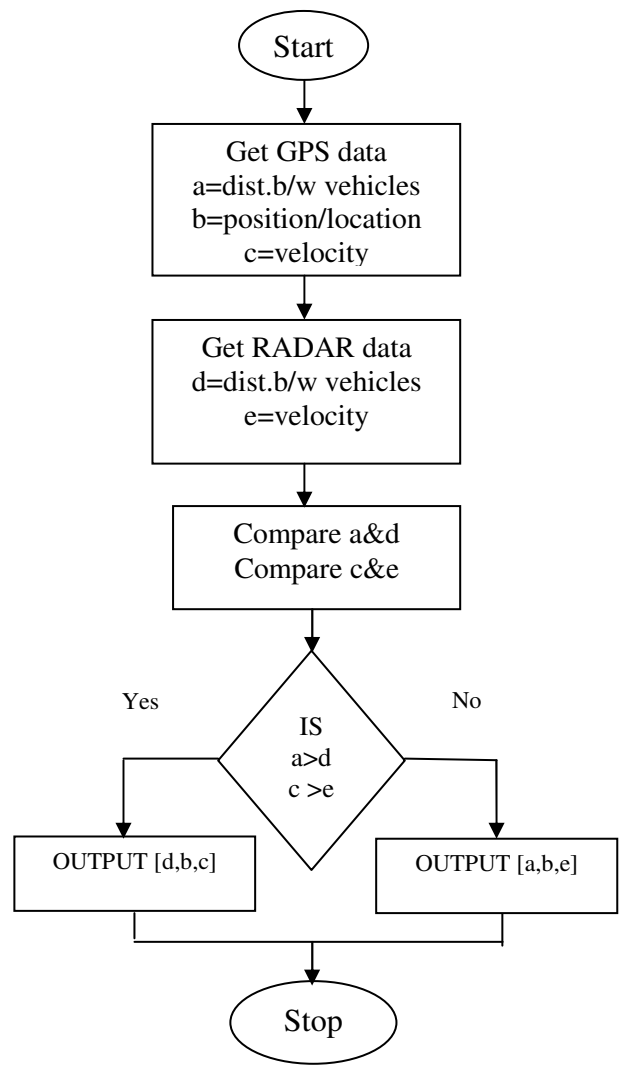

We have used a random source to test our model. The random source is then passed through a channel to make some distortion in it. To recover the data from noisy environment we have used kalman filter. The data which is recovered is very much closed to the original one. This process is followed just to check the efficiency of kalman filter. There are two output of kalman filter (Fig.7 and Fig.8), one is estimated one and other is the predicted one. The estimated data is more reliable than that of predicted data as predicted data comprises of noise and that the estimated data is data, recovered from noisy environment. Though it is not the same as that of original data but is very much close to the original one as shown in Fig.9. Now the fusion block compares the data i.e the distance given by GPS and that by

RADAR. The least distance is then forwarded to the inter vehicle link model. Again the velocity given by the GPS as well as by that of RADAR is compared and then the maximum velocity is then forwarded. The reason for the minimum distance and maximum velocity is suppose the distance between two vehicles is 10m and the distance shown is 12m, then there is a chance that the driver may stop around 10-11m,but the actual distance is near to 10m, so there is a chance between the vehicles, but what if the distance shown is 10m, the driver will definitely stop the vehicle at a distance less than 10m. Though it may be less than the actual distance but the goal of saving life is achieved as collision is avoided.

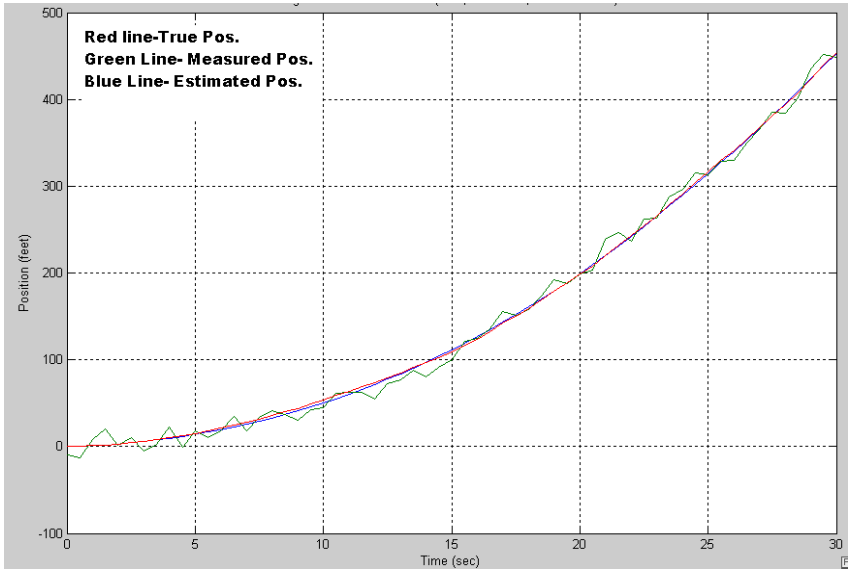

Fig.7. Output of kalman filter which is showing Vehicle position (True, Measured, and Estimated) of moving object.

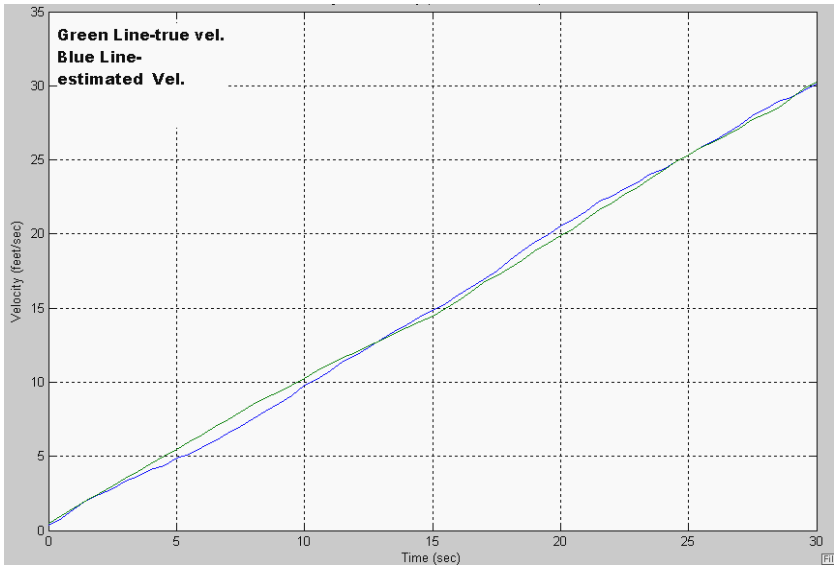

Fig.8. Output of kalman filter which is showing Vehicle velocity (True, and Estimated) of moving object.

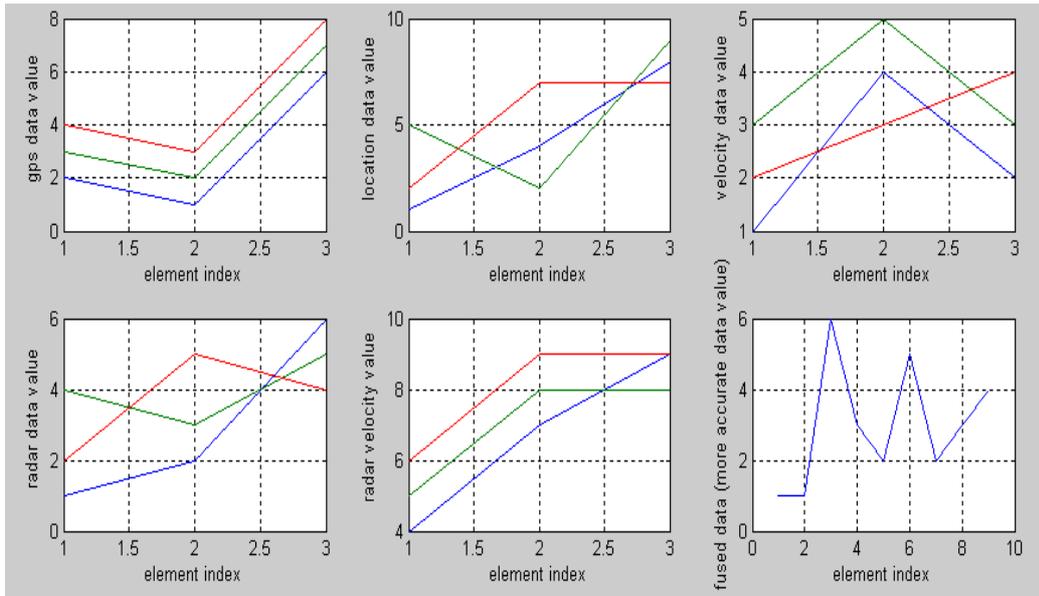

Fig.9 Estimation of GPS ,RADAR and fused data for position ,velocity and location determination.

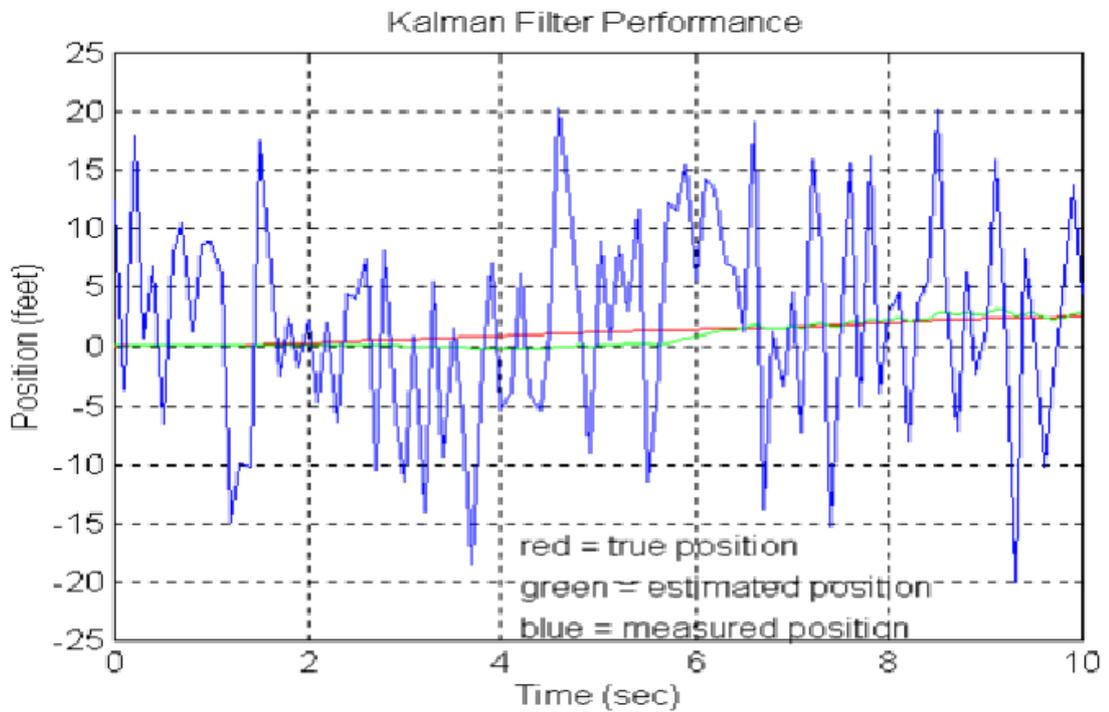

Fig.10. Performance of kalman filter

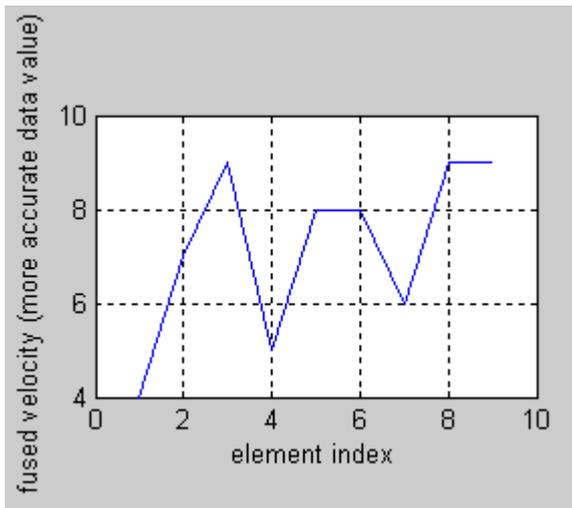

Fig.11. Output of the block diagram.

Fig.10 and Fig .11 depicted the Performance of kalman filter and output of the ITWS.

## CONCLUSION

The objective of this paper is to design a vehicular safety communication architecture that has already been explained. Safety communication to implement ITS has to deal with two principal problems. First, a deterministic channel model cannot be assumed in the fast changing topology of a Vehicular ad-hoc network with its rapidly moving nodes located quite far apart. And Second, hidden terminal collisions cannot be avoided efficiently due to the unbounded system and the broadcast nature of the safety communication. The occurrence of most traffic accidents is based on the simple fact that two or more vehicles are at the same place at the same time. Usually this happens based on an event, such as a hard breaking vehicle, but collisions occur as well if all vehicles are driving in a normal and predictable manner. It is therefore necessary to send so-called routine messages on a regular basis whose information allows the prediction of the vehicle's position for the next few seconds. Whenever this prediction is jeopardized the vehicle might be accelerating or changing its trajectory vigorously a so-called event message has to be distributed in a reliable and timely manner based on an unreliable channel. It is through this distribution of messages that will hopefully reduce the likelihood of accidents. It should be noted that events are generally long lasting since human beings do not operate the vehicle in a very fast manner. This paper presents a new active safety system for automotives. The method presented in this paper extends the situation definition and proposes a design of the actions to have continuity among them. Based on information from GNSS signals, GIS database and local proximity sensors the system is able to produce warning messages in order to avoid possible collisions.To obtain sensorial fusion Multi-Sensor Data Fusion (MSDF) approach is described .These techniques vary from those based on well established Kalman filtering methods, to those based on recent ideas from soft computing technology.

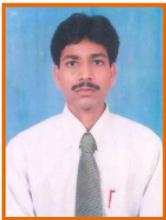

**Prof. Nirmalendu Bikas Sinha** received the B.Sc (Honours in Physics), B. Tech, M. Tech degrees in Radio-Physics and Electronics from Calcutta University, Calcutta,India,in1996,1999 and 2001, respectively. He is currently working towards the Ph.D degree in Electronics and Telecommunication Engineering at BESU. Since 2003, he has been associated with the College of Engineering and Management, Kolaghat. W.B, India where he is currently an Asst.Professor is with the department of Electronics & Communication Engineering & Electronics & Instrumentation Engineering. His current research Interests are in the area of signal processing for high-speed digital communications, signal detection, MIMO, multiuser communications,Microwave /Millimeter wave based Broadband Wireless Mobile Communication ,semiconductor Devices, Remote Sensing, Digital Radar, RCS Imaging, and Wireless 4G communication. He has published large number of papers in different international Conference, proceedings and journals.He is presently the editor and reviewers in different international journals.

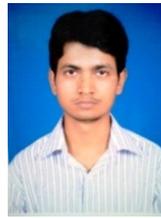

**Manish Sonal** is pursuing B.Tech in the Department of Electronics & Communication Engineering at College of Engineering and Management, Kolaghat, under WBUT in 2011, West Bengal, India..His areas of interest are in Microwave /Millimeter wave based Broadband Wireless Mobile Communication and digital electronics. He has multiple international publications.

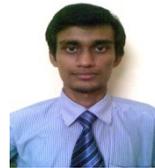

**Makar Chand Snai** is pursuing B.Tech in the Department of Electronics & Communication Engineering at College of Engineering and Management, Kolaghat, under WBUT in 2011, West Bengal, India. His areas of interest are in Microwave /Millimeter wave based Broadband Wireless Mobile Communication and digital electronics. He has published some papers in different international journals.

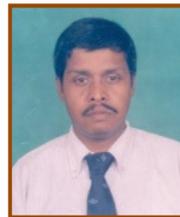

**Dr. Rabindranath Bera** is a professor and Dean (R&D), HOD in Sikkim Manipal University and Ex-reader of Calcutta University, India. B.Tech, M.Tech and Ph.D.degrees from Institute of Radio-Physics and Electronics, Calcutta University. His research areas are in the field of Digital Radar, RCS Imaging, Wireless 4G Communications, Radiometric remote sensing. He has published large number of papers in different national and international Conference and journals.

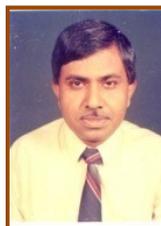

**Dr. Monojit Mitra** is an Assistant Professor in the Department of Electronics & Telecommunication Engineering of Bengal Engineering & Science University, Shibpur. He obtained his B.Tech, M.Tech & Ph. D .degrees from Calcutta University. His research areas are in the field of Microwave & Microelectronics, especially in the fabrication of high frequency solid state devices like IMPATT. He has published large number of papers in different national and international journals. He has handled sponsored research projects of DOE and DRDO. He is a member of IETE (I) and Institution of Engineers (I)society.